\documentclass[twocolumn,amsmath,floatfix,prb,aps]{revtex4-2}

\usepackage{color}
\usepackage{lipsum,amsmath}
\usepackage{pifont}   
\usepackage{graphicx} 
\usepackage{dcolumn}  
\usepackage{bm}       
\usepackage{amsfonts} 
\usepackage{amssymb}  
\usepackage{multirow} 
\usepackage{natbib}
\usepackage{tikz}
\usepackage{graphicx}
\usepackage{subfigure}

\begin{document}
\newcommand{\ba}{{\bf a}}
\newcommand{\BB}{{\bf b}}
\newcommand{\bd}{{\bf d}}
\newcommand{\br}{{\bf r}}
\newcommand{\bp}{{\bf p}}
\newcommand{\bk}{{\bf k}}
\newcommand{\bg}{{\bf g}}
\newcommand{\bt}{{\bf t}}
\newcommand{\bu}{{\bf u}}
\newcommand{\bq}{{\bf q}}
\newcommand{\bG}{{\bf G}}
\newcommand{\bP}{{\bf P}}
\newcommand{\bJ}{{\bf J}}
\newcommand{\bK}{{\bf K}}
\newcommand{\bL}{{\bf L}}
\newcommand{\bR}{{\bf R}}
\newcommand{\bS}{{\bf S}}
\newcommand{\bT}{{\bf T}}
\newcommand{\bQ}{{\bf Q}}
\newcommand{\bA}{{\bf A}}
\newcommand{\bH}{{\bf H}}
\newcommand{\bX}{{\bf X}}
\newcommand{\bsig}{\boldsymbol{\sigma}}
\newcommand{\bdel}{\boldsymbol{\Delta}}

\author{R. Gupta$^{1,2}$}
\email{GUPTARE@tcd.ie}
\author{S. Shallcross$^3$}
\author{J. Quintanilla$^4$}
\author{M. Gradhand$^{1,5}$}
\author{J. Annett$^1$}
\email{James.Annett@bristol.ac.uk}
\affiliation{1 H. H. Wills Physics Laboratory, University of Bristol, Tyndall Ave, BS8-1TL, UK}
\affiliation{2 School of Physics and CRANN, Trinity College, 2, Dublin, Ireland}
\affiliation{3 Max-Born-Institute for non-linear optics, Max-Born Strasse 2A, 12489 Berlin, Germany}
\affiliation{4 Physics of Quantum Materials, School of Physical Sciences,University of Kent, Canterbury CT2 7NH, United Kingdom}
\affiliation{5 Institute of Physics, Johannes Gutenberg University Mainz, Staudingerweg 7, 55128 Mainz, Germany}

\title{Distinguishing $d_{xz}+i d_{yz}$ and $d_{x^2-y^2}$ pairing in $Sr_2RuO_4$ by high magnetic field H-T phase diagrams}

\date{\today}

\begin{abstract}
Employing a realistic tight-binding model describing the Fermi surface in the normal state of $Sr_2RuO_4$ we map out magnetic field versus temperature phase diagrams for $d_{x^2-y^2} (B_{1g})$ and $d_{xz}+id_{yz} (E_g)$ pairing types. Both produce (i) a similar Knight shift suppression of $\sim\!80\%$ and (ii) a bicritical point at $T=0.88$K separating low field second order phase transitions from high field Pauli limiting first order transitions. We find, however, strikingly different phase behaviour within the high field Pauli limiting region. For $d_{x^2-y^2}$ pairing symmetry an additional lower critical line of first order transitions is found (terminating in a critical point at $T=0.09-0.22$K depending on the choice of Hubbard U parameters) while for $d_{xz}+id_{yz}$ no such additional high field phase transitions are found for any choice of Hubbard U. In conjunction with our earlier finding [{\it Physical Review B} {\bf 102} (23), 235203] for $p$-wave helical pairing of a still different high field phase structure (a lower critical field line meeting the upper critical field line exactly at the bicritical point), we suggest high field Pauli limiting phase structure as a possible route to distinguish pairing symmetries in this material.
\end{abstract}

\maketitle

\section{Introduction}

For many years the prevailing view of $Sr_2RuO_4$ was of a well understood Fermi liquid state \cite{Rice-1995} with a phase transition to an unconventional $p$-wave triplet superconductor \cite{science-Nelson, science-Rice, Ishida-1998, Ishida-2001, Luke1998, PhysRevLett.97.167002} of $p_x+ip_y$ order parameter at $T_c = 1.5$K. Recent experiments, reporting significant suppression of the Knight shift below $T_c$ \cite{chronister2020evidence, PhysRevResearch.2.032055, Ishida-2020}, decisively exclude this scenario. The nature of the pairing symmetry in $Sr_2RuO_4$ is thus now the focus of renewed and intense research interest: the parity \cite{chronister2020evidence}, number of components \cite{Ghosh2020, benhabib2020} and even whether the pairing breaks time reversal symmetry (TRS) \cite{Grinenko2021a, Luke1998, PhysRevLett.97.167002, PhysRevB.100.094530} are all the subject of intense discussion amidst conflicting experimental evidence. In such a situation observables that can distinguish between different pairing types are of immense value, and the purpose of the present paper is to suggest a property which it appears has not, to date, been considered as a pathway to discriminate pairing types.

With $p$-wave pairing perhaps ruled out \cite{PhysRevLett.122.027002, PhysRevLett.123.247001, PhysRevResearch.2.032055}, as both chiral or helical pairing types appear not to be able to capture the very strong ($\sim\!80\%$) suppression of the Knight shift \cite{Gupta-2020,Roising-2019}, attention has turned to even parity $d$-wave pairing \cite{Ghosh2020,benhabib2020,Grinenko2021}. The most natural $d$-wave candidates are the TRS breaking chiral $d_{xz} + i d_{yz}$ ($E_{g}$) and TRS preserving $d_{x^2-y^2}$ ($B_{1g}$) order parameters. Discontinuity of elastic constants at $T_c$ \cite{Ghosh2020, benhabib2020} and zero field muon spectroscopy together strongly imply a multi-component order parameter breaking time reversal symmetry, pointing towards $E_{g}$ pairing. However, this would lead to discontinuities in both elastic constants $c_{66}$ and $(c_{11}-c_{12})/2$, with only the former observed \cite{Ghosh2020, benhabib2020}. On the other hand, Bogoliubov quasi-particle interference \cite{Sharma5222} shows clear evidence of vertical line nodes at $k_x=\pm k_y$, and the $E_g$ pairing type has horizontal line nodes in the plane $k_z = 0$. This latter finding in fact supports the $B_{1g}$ pairing type which has exactly such (symmetry imposed) line nodes \cite{Annett-1980}. More exotic multi-component order parameters (${s \pm i d}, {d \pm i g}$) \cite{Ghosh2020, benhabib2020} have been suggested to reconcile these conflicting experiments, but these only produce the observed single superconducting transition as an accidental degeneracy of coupling constants; any change in these, e.g. by compression of the lattice, would then be expected to lead to distinct superconducting transitions for each component, which has not been observed.

One of the most fundamental properties of any material is its phase diagram: the structure of the phase boundaries, the nature of the transitions across them, and how they connect via multicritical points. The temperature-magnetic field phase diagram of $Sr_2RuO_4$ has a well characterized bicritical point at $T=0.8$K and $H=1.2$~Tesla at which the low field second order transition goes over to a first order Pauli limiting transition \cite{Gupta-2020,JPSJ.71.2839, Maeno-2000, Maeno-2000-2, sp-heat-2014, Maeno-2013, mag-2014}. Employing a realistic three-dimensional tight-binding model that very well captures the normal state electronic structure, we explore the H-T phase diagram for the two even parity $d$-wave pairing types $d_{xz}+i d_{yz}$ and $d_{x^2-y^2}$. We find that for both pairing types (i) a similar Knight shift suppression of $\sim\!80\%$ and (ii) a bicritical point of $0.88$K separating lines of first and second transitions. Strikingly, however, these pairing symmetries generate dramatically different phase boundary structures within the high field Pauli limiting region: while $d_{x^2-y^2}$ shows an additional lower critical field line of $1^{\rm st}$ order transitions ending in a critical point at $T=0.09-0.22$K depending on the choice of Hubard $U$ parameters, $d_{xz}+i d_{yz}$ pairing shows no such additional phase boundary for any choice of Hubbard $U$.

In conjunction with our previous finding \cite{Gupta-2020} of a different high field phase diagram structure for helical pairing, the phase structure of the Pauli limiting regime would appear to represent a path to distinguish pairing symmetries in this material. While such upper and lower critical field phase lines have been observed in a number of previous experimental works \cite{PhysRevB.93.184513, JPSJ.71.2839}, they have been found to be sensitive to disorder, with lower mean field path (i.e. lower quality) samples not showing the additional lower critical field seen in higher quality samples \cite{PhysRevB.93.184513}. Nevertheless, in light of our findings and the current uncertainty of the pairing symmetry of this material, experiments to conclusively establish the high field phase diagram could definitively rule out potential candidate pairing states.

\section{Tight-binding Model}
\label{TB}

A realistic 3-dimensional tight binding (TB) model \cite{Annett-2003} was previously used to study the $p$-wave chiral \cite{Gradhand-2013} and helical pairings \cite{Gupta-2020} in $Sr_2RuO_4$. Our effective pairing Hamiltonian is a multi-band attractive $U$ Hubbard model with an "off-site" pairing \cite{Annett-2003}

\begin{eqnarray}
\label{Hamiltonian}
\hat{H} & = & \sum_{ijmm'\sigma}((\varepsilon_m-\mu)\delta_{ij}\delta_{mm'}-t_{mm'}(ij))c^{\dagger}_{im\sigma}c_{jm'\sigma}\nonumber\\
&-&\frac{1}{2}\sum_{ijmm'\sigma\sigma'}U^{\sigma\sigma'}_{mm'}(ij)\hat{n}_{im\sigma}\hat{n}_{jm'\sigma'} +\hat{H}_{so}\nonumber\\
& - & i\lambda\sum_{i,\sigma,\sigma'}\sum_{m,m'}\varepsilon^{\kappa m m'}\sigma^{\kappa}_{\sigma \sigma'}c^\dagger_{i m \sigma}c_{i m' \sigma'};
\end{eqnarray}
where $m$ and $m'$ stand for the three Ruthenium $t_{2g}$ orbitals $a = d_{xy}, b = d_{xz}, c = d_{yz}$ and $i$, $j$ refer to the sites of a body centered tetragonal lattice. Also, we include spin-orbit coupling (SOC) in our model (the last term) where $\sigma^\kappa_{\sigma\sigma'}, \kappa=x,y,z$ are the Pauli matrices, $\varepsilon^{\kappa m m'}$ denotes the completely antisymmetric tensor, and the sign convention implies that the Ru orbital indices must be ordered as $m = (a, b, c)$ in our notation. $\lambda$ represents the strength of the SOC chosen as $12.5$meV \cite{Gradhand-2017} here. The hopping integrals $t_{mm'}(ij)$ and on-site energies $\varepsilon_m$ have been reported in Ref.~[\onlinecite {Annett-2003}], which were fitted to reproduce the experimentally determined FS \cite{Bergemann-2000}, that consists of 3 sheets: $\alpha$ and $\beta$ dominated by $d_{xz}$ and $d_{yz}$ character orbitals, and a $\Gamma$ centred sheet (denoted $\gamma$) dominated by $d_{xy}$ character. The off-site pairing interaction involves two interaction constants $U_{a}$ and $U_{b/c}$.

\subsection{Pairing function}
\label{pairing}

As $Sr_2RuO_4$ has a high symmetry body-centered tetragonal crystal structure there exist many symmetry allowed choices of the pairing function, corresponding to different irreducible representations of the point group symmetry \cite{Annett-1980}. In this work we consider two most probable $d$-wave pairings in $Sr_2RuO_4$: $d_{xz}+i d_{yz}$ ($E_g$ irreducible representation) which is the simplest d-wave state with TRS breaking and $d_{x^2-y^2}$ ($B_{1g}$ irreducible representation) which has no TRS breaking, but would be expected from some spin fluctuation pairing models \cite{PhysRevLett.123.247001}. 

For the $B_{1g}$ model the most natural even-parity basis functions are given by

\begin{align}
\label{basis1}
\cos k_x, \,\,\cos k_y 
\end{align}

while for the $E_g$ model, these are

\begin{align}
\label{basis2}
\sin \frac{k_x}{2}\cos \frac{k_y}{2}\sin\frac{k_z c}{2},\,\, \cos \frac{k_x}{2}\sin \frac{k_y}{2}\sin\frac{k_z c}{2}
\end{align}
which are the simplest basis functions having the symmetry required node in the $k_z=0$ plane \cite{INS-2020}.
In addition, for the $B_{1g}$ model we consider only in-plane nearest neighbour pairing for all the three orbitals: In this model the symmetry of the pairing function does not allow coupling between the adjacent planes and therefore we consider a 2D pairing model. On the contrary, for $E_g$ pairing, symmetry prohibits in-plane pairing and we consider coupling only between the adjacent planes. In general, the pairing can be written as a $3\times3$ matrix in orbital space as:

\begin{align}
\label{U-matrix}
U_{m,m'}=\begin{pmatrix}
U_a & 0 & 0\\
0 & U_{b/c} & U_{b/c}\\
0 & U_{b/c} & U_{b/c}
\end{pmatrix}
\end{align}
with the only difference for the two models being that the U-parameters represent in-plane pairing for the $B_{1g}$ model whereas they correspond to out-of-plane pairing for the $E_g$ model. 

Following the approach of Ref.~[\onlinecite{Gradhand-2013}] and using the basis functions \eqref{basis1} and \eqref{basis2}, the gap function can be written for $E_{g}$ pairing as

\begin{align}
\label{GF1}
\Delta_{ij}(\bk)=\biggl(\Delta^{\uparrow\downarrow,x}_{ij}\sin \frac{k_x }{2}\cos \frac{k_y}{2}\nonumber\\
+i\Delta^{\downarrow\uparrow,y}_{ij}\cos \frac{k_x }{2}\sin \frac{k_y}{2}\biggr)\sin\frac{k_z c}{2},
\end{align}
and for $B_{1g}$ pairing as
\begin{align}
\label{GF2}
\Delta_{ij}(\bk)=\biggl(\Delta^{\uparrow\downarrow,x}_{ij}\cos k_x-\Delta^{\downarrow\uparrow,y}_{ij}\cos k_y\biggr).
\end{align}
The coefficients involved are given by

\begin{align}
\label{C1}
\Delta^{\sigma\sigma',x}_{ij}=4U\sum_n\int d^3k ~u^\sigma_{i,n}(\bk) v^{\sigma'\star}_{j,n}(\bk)\sin \frac{k_x }{2}\nonumber\\
\times\cos \frac{k_y}{2}\sin\frac{k_z c}{2}(1-2f(T,E_n)),
\end{align}
for the $E_{g}$ model and by
\begin{align}
\label{C2}
\Delta^{\sigma\sigma',x}_{ij}=U\sum_n\int d^3k ~u^\sigma_{i,n}(\bk) v^{\sigma'\star}_{j,n}(\bk)\cos k_x \nonumber\\
\times(1-2f(T,E_n)),
\end{align}
for the $B_{1g}$ pairing. Similar relations hold for the $y$-components $\Delta^{\sigma\sigma',y}_{ij}$. In these expressions $f(T,E_n)$ is the Fermi function at a temperature $T$ and eigenvalue $E_n$ corresponding to the $n^{th}$ band. $U=U_a$ for $a-a$ pairing and $U=U_{b/c}$ otherwise. Also, we include interorbital coupling between $b$ and $c$ orbitals in our model though it is extremely weak.  

Using the above equations, along with the symmetry relations \cite{Gradhand-2013}
\begin{eqnarray}
\Delta^{\sigma\sigma',x}_{aa} & = &\Delta^{\sigma\sigma',y}_{aa} \label{sym1}\\
\Delta^{\sigma\sigma', x/y}_{bb} & = &\Delta^{\sigma\sigma', y/x}_{cc}\nonumber,
\label{sym2}
\end{eqnarray}
we solve the Bogoliubov de Gennes (BdG) equation

\begin{align}
\label{BdG}
\begin{pmatrix}
\hat{H}_{\bk}(\br) & \hat{\Delta}_{\bk}(\br)\\
\hat{\Delta}^{\dagger}_{\bk}(\br) & -\hat{H}^*_{-\bk}(\br)
\end{pmatrix}
\begin{pmatrix}
u_{n\bk}(\br)\\
v_{n\bk}(\br)
\end{pmatrix}=E_{n\bk}
\begin{pmatrix}
u_{n\bk}(\br)\\
v_{n\bk}(\br)
\end{pmatrix},
\end{align}
at every $k$-point of a $480\times480\times48$ mesh. 
The only unknown constants are the interaction parameters $U_{a}$ and $U_{b/c}$. These are chosen such that there is a single superconducting critical temperature of $1.5$K. Under this requirement we find

\begin{eqnarray}
U_a&=&0.3107\,t \\
U_{b/c}&=&0.618\,t\nonumber
\end{eqnarray}
for $E_{g}$ pairing and 

\begin{eqnarray}
U_a&=&0.2352\,t \\
U_{b/c}&=&0.9705\,t,\nonumber
\end{eqnarray}
for $B_{1g}$ pairing with $t=0.08162$~eV. It should be noted that the presence of SOC in our model couples the orbitals together. Therefore, it is a realistic model where the system is more likely to have a common superconducting transition.

In our TB model a spin-only magnetic field $\bH=(H_x,H_y,H_z)$ can be added to Eq.~\eqref{BdG} by replacing $\hat{H}_{\bk}(\br)$ with

\begin{align}
\hat{H}_{\bk}(\br)={H}_{\bk}(\br)\hat{\sigma}_0+\mu_B \mu_0 \hat{\bsig}.H,
\label{spiny}
\end{align}
with $\mu_B$ the Bohr magneton and $\mu_0$ the vacuum permeability (in what follows we set $\mu_0=1$ for convenience).

\section{Zero-field properties}

We will first explore the zero and low external field properties of $Sr_2RuO_4$ for the two pairing symmetries we consider: the gap structure on the Fermi surface, density of states, and zero field specific heat. For each case we find that the interaction parameters chosen above lead to good agreement with experiment and, moreover, comparably good agreement for both $d_{x^2-y^2}$ and $d_{xz}+id_{yz}$ pairing.

\begin{figure*}
\includegraphics[width=.98\linewidth]{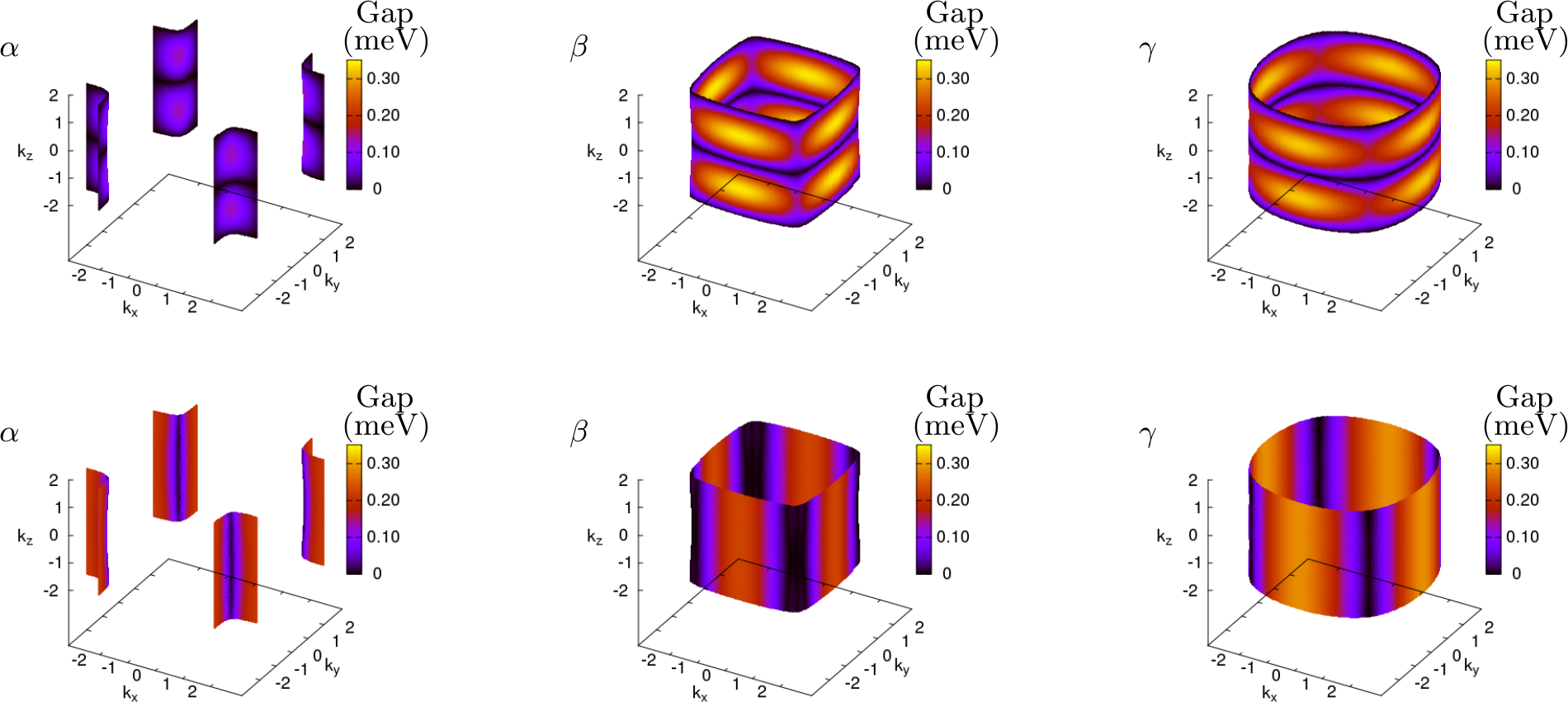}
\caption{{(Color online)} Variation of zero temperature superconducting gap on the three Fermi sheets of the normal state of $Sr_2RuO_4$. The top (bottom) row is for the $E_g$ ($B_{1g}$) pairing. Whereas the $E_g$ pairing has horizontal line nodes at $k_z=0,\pm 2\pi/c$, $B_{1g}$ has vertical line nodes at $k_x=\pm k_y$. $k_x, k_y$ and $k_z$ are in units of the in-plane lattice constant $a=3.862$~{\text\AA}. $c=12.722$~\text\AA  ~is the lattice constant along $z$-axis.}
\label{FS}
\end{figure*}

{\it Gap structure}: In Fig.~\ref{FS} we show the superconducting gap $\Delta(\bk)$ obtained by solving the BdG equation (Eq.~\eqref{BdG}). As can be seen from the scale bars, the gap varies between $0$meV and $\sim 0.35$meV. Interestingly, $\sim$ $0.35$meV is also the maximum value of the single-particle superconducting gap reported in the Bogoliubov quasiparticle interference (BQPI) measurements \cite{Sharma5222} and also from differential conductance spectra \cite{DOS2,DOS1}. The presence of nodes in the gap function can clearly be observed: as expected, these are horizontal for $d_{xz}+id_{yz}$ pairing and vertical for $d_{x^2-y^2}$ pairing and are found on each of the $\alpha$, $\beta$, and $\gamma$ Fermi sheets. The positions of these nodal lines can be readily understood from the gap functions in Eqs.~\eqref{GF1} and \eqref{GF2}: for $d_{xz}+id_{yz}$ pairing we have $\sin(k_z c/2)=0$ yielding horizontal nodal lines at $k_z=0$ and $k_z=\pm 2\pi/c$, while for $d_{x^2-y^2}$ pairing the corresponding condition is $\cos k_x=\cos k_y$, yielding vertical nodes when $k_x=\pm k_y$. These nodes contrast with the chiral and helical $p$-wave \cite{Gupta-2020} scenarios which have horizontal line nodes (though not symmetry imposed) on the $\alpha$ and $\beta$ sheets for $k_z=\pm \pi/c$ planes while the $\gamma$ sheet has deep minima. Experiments on determining the nodal structure of the gap function have conflicting results: vertical line nodes are supported by thermal conductivity measurements and BQPI measurements \cite{Sharma5222}, with horizontal line nodes by spin resonance in inelastic neutron scattering measurements \cite{INS-2020} and specific heat capacity measurements \cite{Field-dependent}.

\begin{figure}
\includegraphics[width=.98\linewidth]{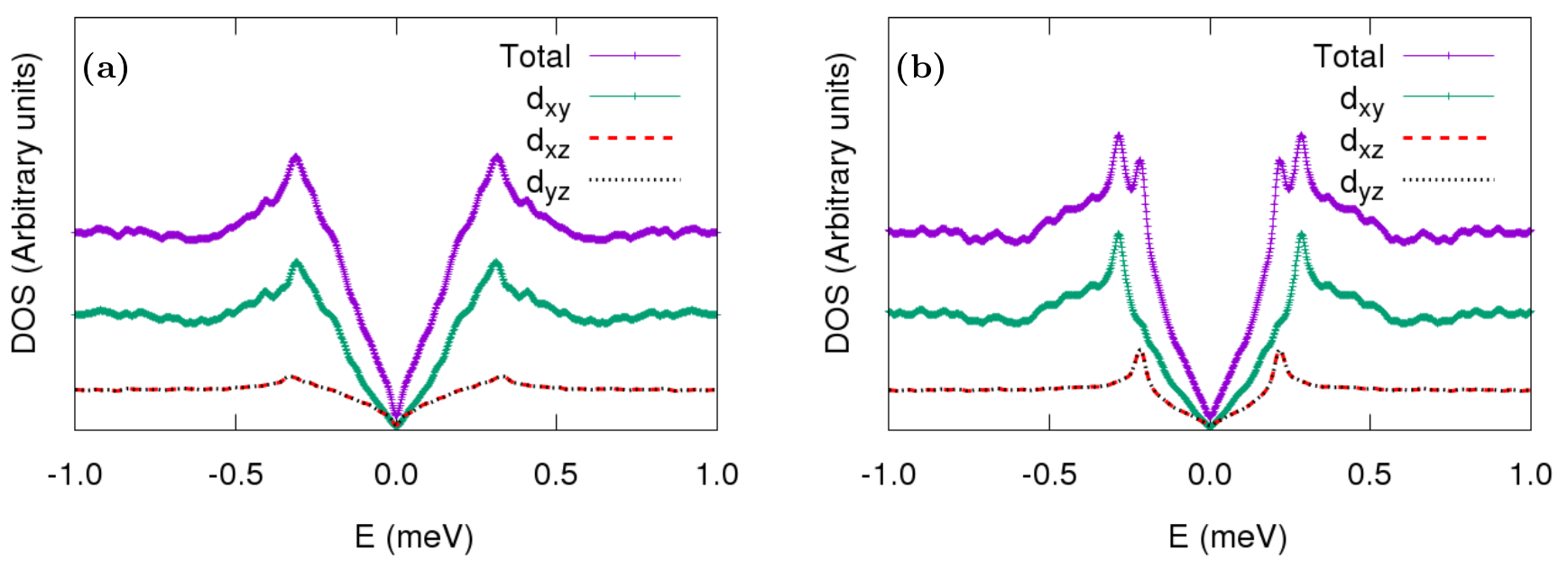}
\caption{{(Color online)} Orbital resolved and total superconducting density of states (DOS) for the a) $E_g$ and b) $B_{1g}$ model. Linear DOS at low energies is a consequence of the presence of nodal lines. Normal-state contribution to the total DOS from $d_{xy}, d_{yz}$ and, $d_{xz}$ orbitals is $\sim 58\%, 21\%$ and $21\%$ respectively. The superconducting gap 2$\Delta$ from the separation between the peaks is $\sim 0.56$meV. At low energies, $E_g$ pairing has a larger total DOS as compared to the $B_{1g}$ pairing. Contribution from $d_{xy}$ orbital dominates at nearly all energies in both the models.}
\label{DOS}
\end{figure}

{\it Density of states}: In Fig.~\ref{DOS}, we show the orbital resolved and total superconducting density of states (DOS) for the two pairing functions. As expected, due to the presence of line nodes the DOS is linear close to zero energy. The contribution to the total DOS from $d_{xy}, d_{yz}$ and, $d_{xz}$ orbitals is $\sim 58\%, 21\%$ and $21\%$ respectively in the normal state. The ratio for the orbital contribution $d_{yz}+d_{xz}:d_{xy}= 42:58$ matches well with the ratio of $43:57$ for the contribution from $\alpha+\beta$ and $\gamma$ bands from de Haas-van Alphen measurements \cite{de-Haas}. The value of the superconducting gap 2$\Delta$ (the separation between the peaks in Fig.~\ref{DOS}) is $\sim 0.56$meV, closely agreeing with tunneling spectroscopy measurements \cite{DOS2,DOS1}.
 
\begin{figure}
\includegraphics[width=.98\linewidth]{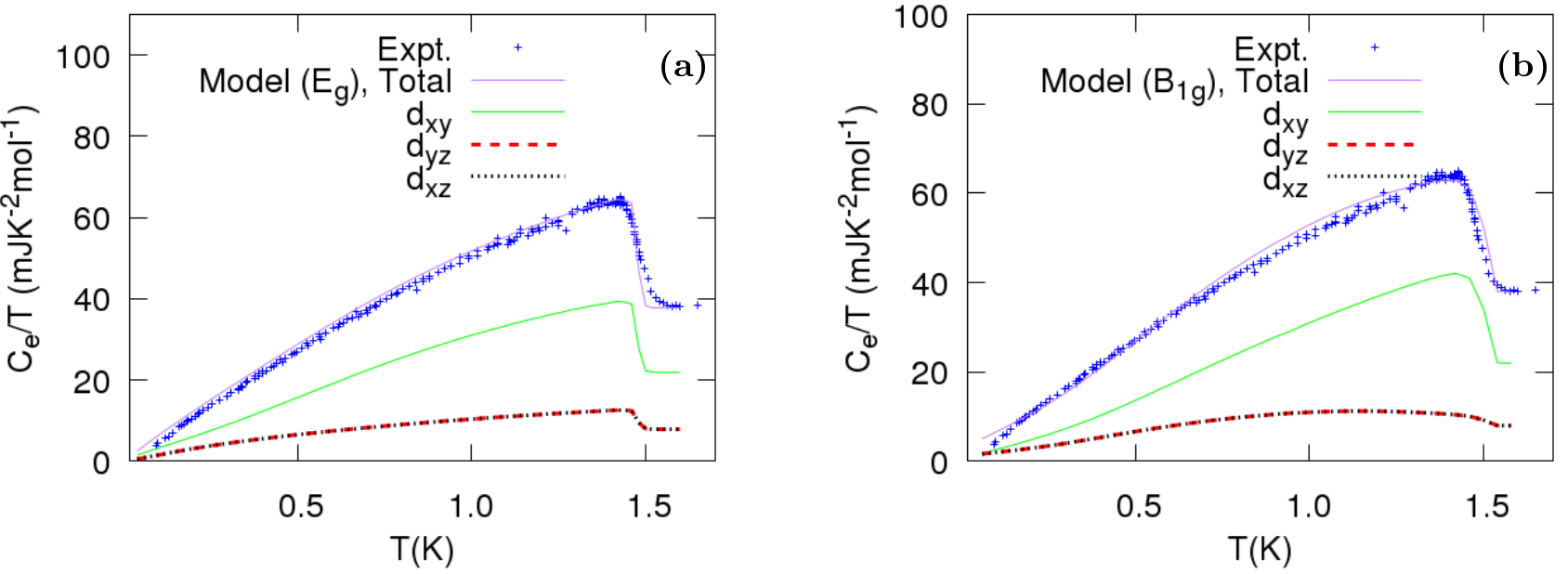}
\caption{{(Color online)} Temperature variation of zero-field specific heat for a) $E_g$ and b) $B_{1g}$ pairing. Comparison with the experimental data \cite{Maeno-2000} shows a good agreement for both the models. Linear behavior at low temperatures is due to the presence of line nodes. Overall, we see a somewhat better agreement for $E_g$ pairing compared to the $B_{1g}$ pairing.}
\label{CV-0}
\end{figure}

{\it Specific heat}: Correctly capturing the zero field specific heat represents an important test of any theory of $Sr_2RuO_4$. In Fig.~\ref{CV-0} we show the orbital resolved and total specific heat for the two pairing types we consider. As can be seen a good agreement with the experimental specific heat \cite{Maeno-2000} exists for both $E_g$ pairing and $B_{1g}$, including for the magnitude of the superconducting jump. Note that the low-energy linear behavior with temperature is due to the presence of line nodes on the Fermi surface. The ratio of the contribution from $d_{xy}$, $d_{yz} + d_{xz}$ orbitals for the $E_g$ and $B_{1g}$ models is $64:36$ and $82:18$ respectively.

\section{Properties at finite field} 

The physics of $Sr_2RuO_4$ in a magnetic field offers key insights into the pairing symmetry of the superconducting order parameter; notably the Knight shift for a-b plane oriented magnetic fields for which (in the absence of Fermi liquid corrections and SOC) theory predicts at $T=0$ a $50\%$ decrease in its value for helical $p$-wave pairing, no decrease for chiral $p$-wave pairing, and a $100\%$ drop for $d$-wave pairing \cite{Annett-1980}, \cite{Legget} under the application of a weak in-plane magnetic field. Here, we will first explore the spin polarization (in the normal and superconducting state), spin susceptibility, and specific heat, before from the latter property constructing the $H-T$ phase diagram of $d_{xz}+i d_{yz}$ and $d_{x^2-y^2}$ pairing in $Sr_2RuO_4$. 

One should note that that our magnetic field couples only with the spin degree of freedom, see Eq.~\ref{spiny} (although the presence of SOC can induce a small orbital magnetization), and so the vortex lattice contribution to the physics has been ignored. Quantitative agreement with critical magnetic fields and values of magnetic moments of experiments is thus not to be expected.

\subsection{Spin-moments and spin susceptibility}

\begin{figure}
\centering
\begin{tikzpicture}
\draw (0, 0) node[inner sep=0]
{\includegraphics[width=0.9\linewidth]{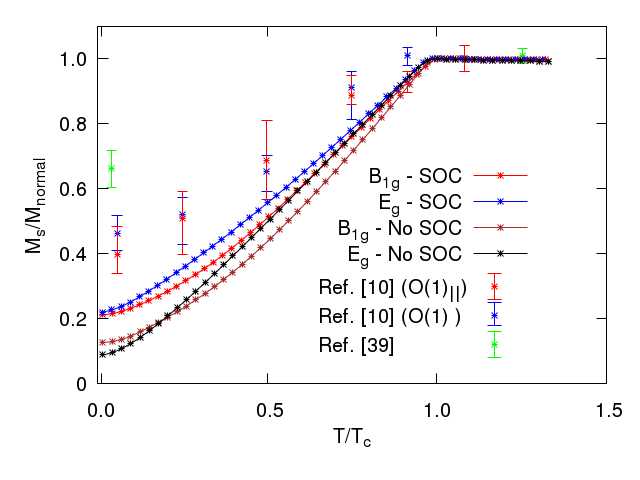}};
\draw (1.62,-1.03) node {$_\perp$};
\end{tikzpicture}
\caption{{(Color online)} Temperature variation of the ratio of spin moment in the superconducting and normal state at $\sim 0.7$T for $d_{xz}+i d_{yz}$ ($E_g$) and $d_{x^2-y^2}$ ($B_{1g}$) pairing. Shown also is the same data with spin orbit coupling (SOC) switched off. Experimental data for the Knight shift at $0.7$T from NMR measurements and PNS data at $0.5$T are also shown for comparison.}
\label{KS-T2}
\end{figure}

{\it Spin moments}: In Fig.~\ref{KS-T2} we present the ratio of the spin polarization in the superconducting and normal state ($M_s/M_{normal}$) as a function of temperature with the magnetic field held fixed. We set the field to $\sim 0.7$~Tesla in order to compare with the experimental data available from NMR measurements \cite{Ishida-2020}. Also displayed in this figure is the $0.5$T data from PNS measurements \cite{Alex-2020}. The zero-temperature magnetization has a drop of $\sim 80\%$ for both the $d$-wave pairing models; turning off spin orbit coupling in the calculation reduces this further to $\sim 90\%$. The ratio of $M_s/M_{normal}$ is very close for both pairing types; for the $E_g$ pairing which breaks TRS we also find a small orbital component \cite{Gradhand-2020}. The absence of the flux lattice contribution in particular means that quantitative agreement with PNS data at $0.06$K is not expected, as magnetic neutron diffraction couples to the total magnetic response (spin, orbital, and diamagnetic) with each component weighted equally.

\begin{figure}
\centering
\begin{tikzpicture}
\draw (0, 0) node[inner sep=0] {\includegraphics[width=0.9\linewidth]{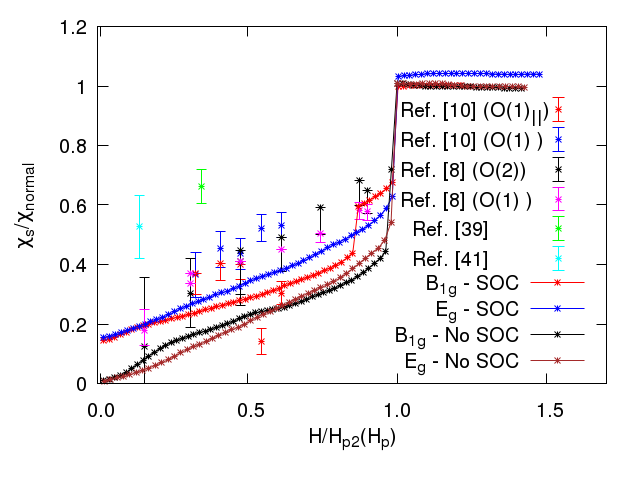}};
\draw (2.6, 1.13) node {$_\perp$};
\draw (2.47, 0.39) node {$_\perp$};
\end{tikzpicture}
\caption{{(Color online)} Spin susceptibility ratio as a function of field for temperature value of $25$mK for $d_{xz}+i d_{yz}$ ($E_g$) and $d_{x^2-y^2}$ ($B_{1g}$) pairing. Also presented is experimental data for the Knight shift measured at two different positions of Oxygen atoms at $66$mK \cite{Ishida-2020} and at $25$mK \cite{chronister2020evidence} and the PNS data at $60$mK \cite{Alex-2020} and at $25$mK \cite{Alex-unpublished}. Note that the applied magnetic field has been divided by the critical field value $H_{p}$, which for the case of $d_{x^2-y^2}$ is chosen to be the upper critical field $H_{p2}$, see the discussion of the $H-T$ phase diagram in Sec.~\ref{H-T}.}
\label{KS-T}
\end{figure}

{\it Spin susceptibility}: In Fig.~\ref{KS-T} we show the spin susceptibility ratio between the superconducting and the normal states as a function of field for $T=25$mK with the applied magnetic field scaled by the upper critical field. Also shown is the experimental data for the Knight shift obtained by NMR at two different oxygen positions (for O(1) at $66$mK \cite{Ishida-2020} and for O(2) at $25$mK \cite{chronister2020evidence}), along with neutron scattering data measured at $60$mK \cite{Alex-2020} and $25$mK \cite{Alex-unpublished}. Within the broad scatter of experimental points, it can be seen that both pairing types agree well with the data. The magnitude of the Knight shift suppression therefore does not represent a quantity capable of discriminating between these $d$-wave pairing types. However, a striking difference can be seen: the susceptibility ratio for $d_{{x^2}-{y^2}}$ shows an addition jump after the critical field at which the superconducting transition occurs, indicating a second phase transition. Evidently the large error bars preclude identification of this feature from the experimental data, and so we now turn to other signatures of this additional transition.

\subsection{Specific heat}

\begin{figure}
\includegraphics[width=.98\linewidth]{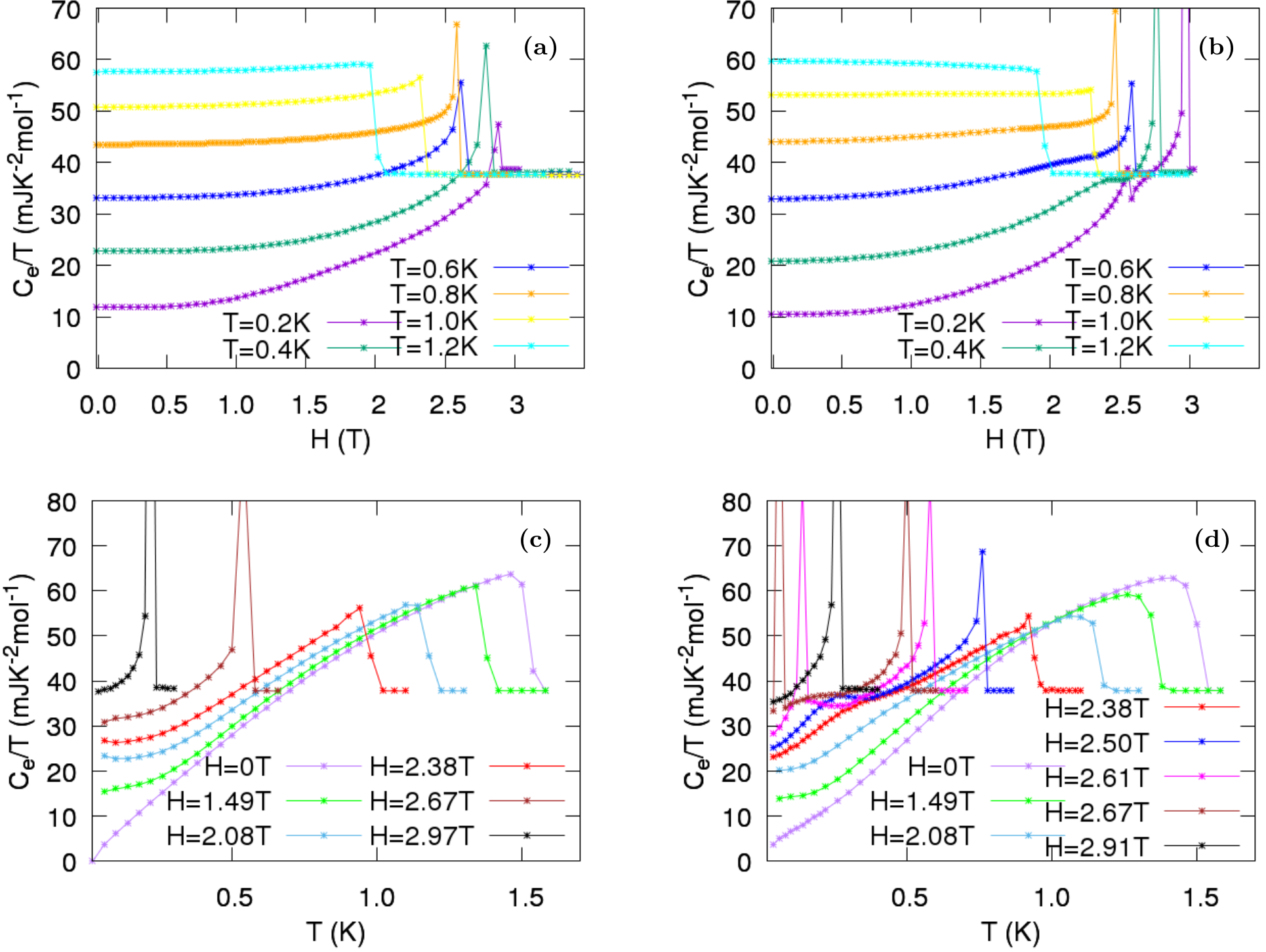}
\caption{{(Color online)} Magnetic field variation of the specific heat for (a) $d_{xz}+id_{yz}$ pairing ($E_g$) and (b) $d_{x^2-y^2}$ pairing ($B_{1g}$) for a series of fixed temperatures. For temperatures less than $T \leq T^*=0.88$K a clear change can be seen from a step like feature to a pronounced peak, marking the onset of Pauli limiting at which the transition goes from second order to first order. Interestingly, the Pauli limiting onset temperature is identical for both pairing types, and agrees well with the experimental value (0.8K). While the heat capacity curves are very similar for both pairing types for $T=0.2$ a clear second peak can be seen for $d_{x^2-y^2}$ indicating a second superconducting transition, which does not occur for $d_{xz}+id_{yz}$ pairing. Similar behaviour is seen in panels (c) and (d), in which is displayed the corresponding behaviour for temperature variation with fixed field. In this case the crossover field to Pauli limiting at $H^*\sim 2.38$T is overestimated as compared to experiment (which finds $H^*\sim 1.2$T), a result of coupling the external field only to spin and not orbital degrees of freedom.}
\label{CV-T-B}
\end{figure}

Presented in Fig.~\ref{CV-T-B}a, b is the specific heat for varying in-plane magnetic field for a series of fixed temperatures between $T=0.2$ and $T=1.2$K. For both $d_{xz}+i d_{yz}$ and $d_{x^2}-{y^2}$ pairing we see the evolution of pronounced peaks at the critical field at low temperatures, to a step like feature at higher temperatures, indicating a crossover from first order to second order transitions at the critical field. This crossover is due to Pauli limiting, and has been observed in several experiments \cite{JPSJ.71.2839, Maeno-2000, Maeno-2000-2, sp-heat-2014, Maeno-2013, mag-2014}. Interestingly, the temperature separating first and second order transitions is found to be the same for both pairing types. Reassuringly, while we fit our model parameters to reproduce the zero field critical temperature, the value we find of $T^*=0.88$K agrees very closely with the experimental value of $T^*=0.80$K for the temperature onset of Pauli limiting.

While $T^*$ is identical for both pairing types, the behaviour within the Pauli limiting low temperature and high field regime is strikingly different. As can be noted from examination of Fig.~\ref{CV-T-B} for $T > T_{c2} = 0.2$K the specific heat curves are nearly identical, while for $T \le T_{c2}$ they become quite different. We see first a distinct shoulder in the specific heat at $T = 0.4$K indicating the onset of a phase instability, which has evidently occurred by $T=0.2$K where an additional peak in the heat capacity can clearly be seen. 

Turning to the heat capacity for varying temperature with the in-plane magnetic field held fixed between $T=0.2$ and $T=1.2$K in Fig~\ref{CV-T-B}c, d, we see (as expected) a similar behaviour. While the heat capacities are very similar for field strengths of up to $H = 2.38T$, for higher fields the $d_{x^2-y^2}$ pairing shows again the development of a shoulder feature (for $H = 2.38T$) going over to a second peak that is clear in the curves for $H = 2.61T$ and $H = 2.67T$. For higher field strengths a single transition is again seen. The field strengths for the onset of Pauli limiting behaviour and the appearance of a second phase transition appear to be nearly identical. As expected, since we couple the field only to spin, the critical field for the onset of Pauli limiting first order transitions is too high as compared to the field found in experiment ($\sim 1.2$~Tesla) \cite{PhysRevB.93.184513}.

\subsection{H-T phase diagrams}
\label{H-T}
\begin{figure}
\includegraphics[width=.98\linewidth]{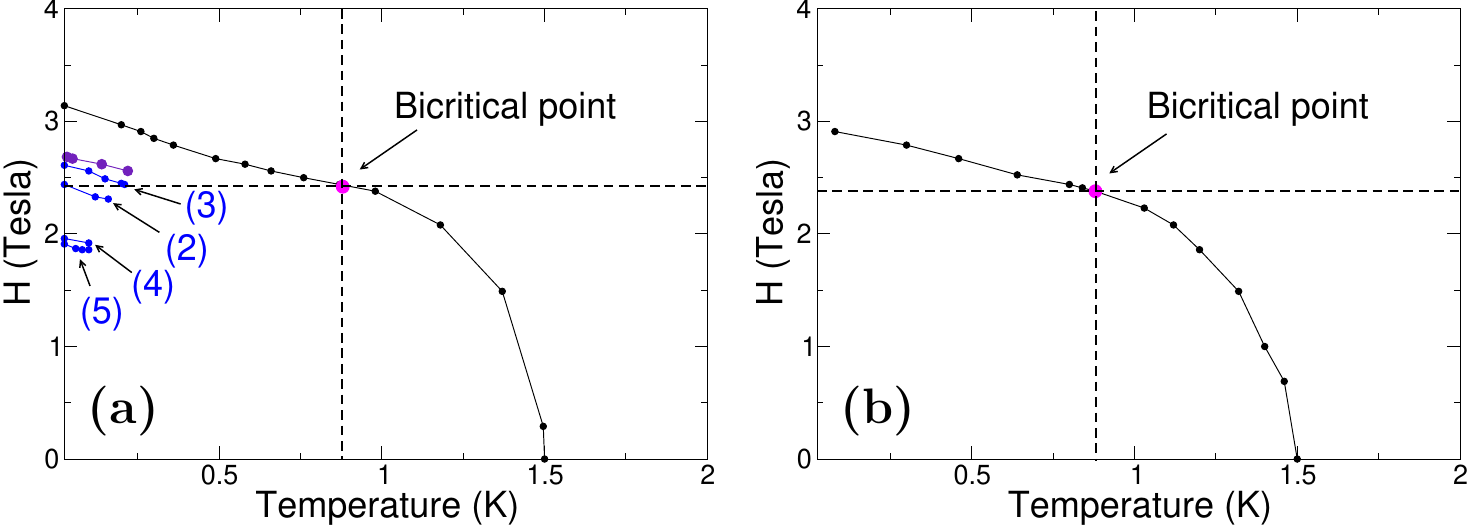}
\caption{{(Color online)} H-T phase diagram for (a) $d_{x^2-y^2}$ pairing ($B_{1g}$) and (b) $d_{xz}+id_{yz}$ ($E_g$). For both pairing types a bicritical point exists marking the onset of Pauli limited first order transitions; we find $T^*=0.88$K and $H^*=2.38$~Tesla, the former in excellent quantitative with experiment, the latter overestimated due to the absence of the vortex lattice in our calculations. Strikingly, in the high field low temperature Pauli limiting regime $d_{x^2-y^2}$ pairing shows an additional line of first order transitions ending in a critical point at $T=0.22$K. The additional phase lines labelled (2-5) represent the variation of this additional phase line upon variation of the Hubbard $U$ parameters of our model; for no choice of Hubbard $U$ are additional phase structures seen for $d_{xz}+id_{yz}$ pairing. (Note that the variation of Hubbard $U$ does not change the upper critical line or bicritical point significantly.)}
\label{phase}
\end{figure}

Finally, from these and data from other temperature and field slices we construct the H-T phase diagrams of these two pairing types. These are shown in Fig.~\ref{phase}. As can be seen, the bicritical point ($T^*, H^*$) separating the Pauli limiting first order from the low field second order transitions is very similar for both pairing types ($T^*=0.88, H^*=2.42$) for $B_{1g}$ and ($T^*=0.88, H^*=2.38$) for $E_g$; in a previous work we found the same values of ($T^*=0.8, H^*=2.5$) for $p$-wave helical pairing \cite{Gupta-2020} indicating that this feature is very robust to the superconducting order parameter. As noted above, while $T^*$ agrees very well with experiment, the absence of a flux lattice in our calculations results in a much higher field value of the bicritical point than the $1.2$~Tesla \cite{PhysRevB.93.184513} found in experiment. The additional transitions found for $d_{x^2-y^2}$ pairing can now clearly be seen as a line of additional first order transitions below the upper critical field, terminating in a critical point (at $T=0.22$K and $H=2.56$~Tesla on line (3) in Fig.~\ref{phase} (a) for our choice of Hubbard parameters). This additional phase behaviour in the Pauli limiting regime represents a striking thermodynamic difference between the two pairing types. Therefore, whereas there is only one critical field ($H_p$) for the $E_g$ pairing, two critical fields ($H_{p1}$ and $H_{p2}$) exist for the $B_{1g}$ pairing (with $H_{p2}$ being only in a certain temperature interval).

\begin{figure*}
\includegraphics[width=.98\linewidth]{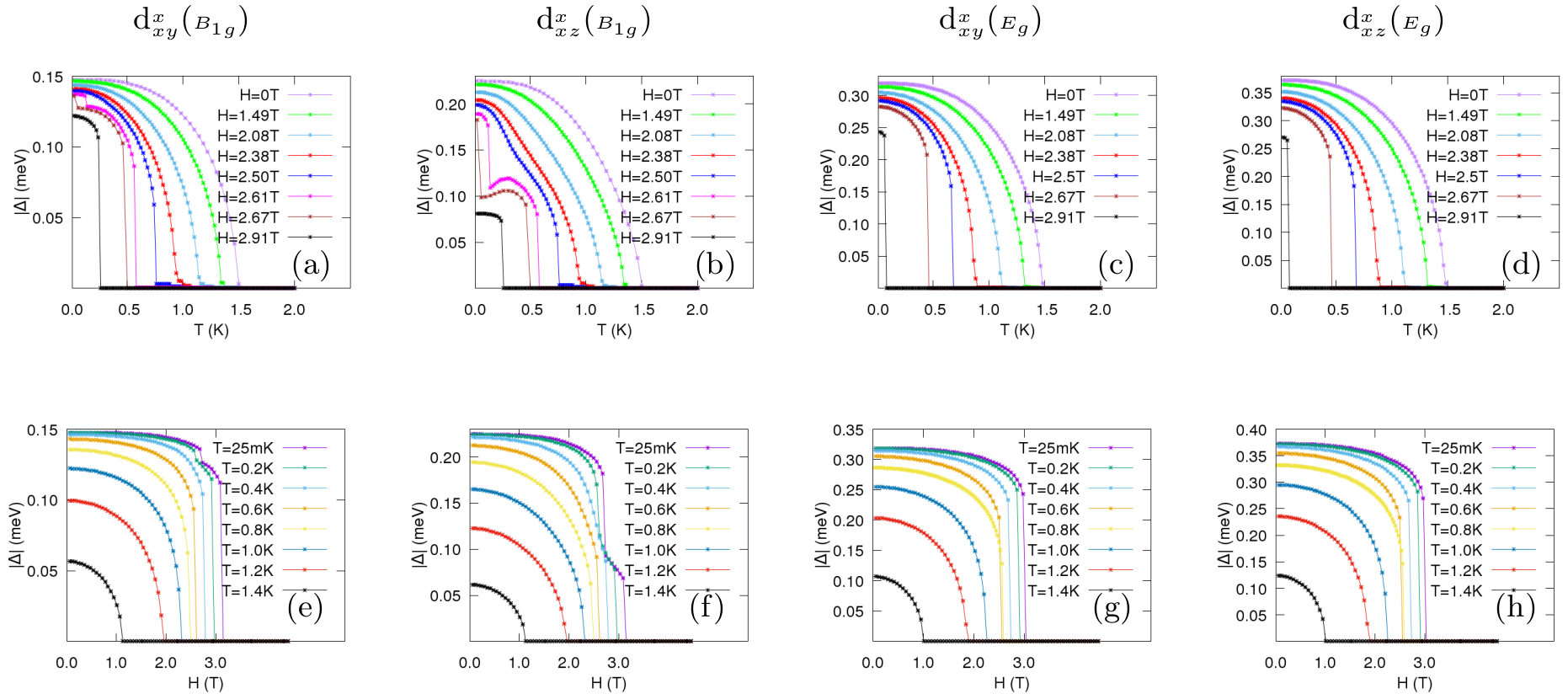}
\caption{{(Color online)} Temperature variation of the $x$-component of gap-function on $d_{xy}$ and $d_{xz}$ orbitals for  $d_{x^2-y^2}$ pairing ($B_{1g}$), panels a and b, and $d_{xz}+id_{yz}$ pairing ($E_g$), panels c and d. Clearly visible are the crossover from second order to first order transitions with the onset of Pauli limiting, and the second additional transition for $d_{x^2-y^2}$ pairing in which a collapse of the gap function by nearly 50\% of its value occurs on the $\alpha$ and $\beta$ Fermi sheets (a small corresponding jump can be seen in $d_{xy}^x$ i.e. on the $\gamma$ Fermi sheet, induced by spin-orbit coupling). Similar behaviour can be seen for field variation at fixed temperature, see panels e-f, $d_{xy}$ and $d_{xz}$ orbitals for $d_{x^2-y^2}$ pairing, and g-h, $d_{xy}$ and $d_{xz}$ orbitals for $d_{xz}+id_{yz}$ pairing.}
\label{Delta-T-B}
\end{figure*}

The appearance of additional phase transitions in the Pauli limiting regime has been observed in, to the best of our knowledge, two previous experiments. In magnetic torque measurements a lower critical line of first order transitions was found in long mean field path samples, but not in (the presumably more disordered) samples of lower mean free path \cite{PhysRevB.93.184513}. As we find here, this additional line of transitions extended from $T=0$ with negative slope and terminating before the bicritical point (at $T=0.7$). A double superconducting transition was observed in the heat capacity both for fixed field and fixed temperature \cite{JPSJ.71.2839}, again indicating a negative slope of the additional line of transitions. 

We should note that sensible variation of the Hubbard $U$ parameters of our model does not qualitatively change these findings. In Fig.~\ref{phase} we show the phase boundary lines obtained by 4 choices of Hubbard U parameters labelled (2)-(5) which are, in order, $(0.2360,0.9500)$, $(0.2300,0.9705)$, $(0.2340,0.9200)$ and $(0.2360,0.9100)$ with the number pair denoting the $U_a$ and $U_{b/c}$ Hubbard $U$ parameters respectively (see, Sec.~IIA) in units of $t=0.08162$eV. Each pair of Hubbard $U$ parameters, which differ only by a few percent from those given in Sec.~IIA and used throughout this work, both reproduce the zero field $T_c$ of 1.5K as well as the Knight shift suppression. However, as can be noted from Fig.~\ref{phase}, these different choices of Hubbard $U$ parameters result in a variation of the critical end point in the range $0.09-0.22$K and a variation in the critical field over nearly 1 Tesla. On the other hand for all choices of Hubbard $U$ parameters the $d_{xz}+id_{yz}$ pairing never shows the additional phase line. We thus conclude that the finding of an additional phase transition in the Pauli limiting region for $d_{x^2-y^2}$ pairing but not $d_{xz}+id_{yz}$ pairing is robust within our model.

\subsection{Gap-function}
Finally, in Fig.~\ref{Delta-T-B} we present the gap function for both fixed field varying temperature and vice versa. We first consider $d_{x^2-y^2}$ pairing, and panels (a,b) present the $x$-component of $\Delta_{xy}$ and $\Delta_{xz}$ respectively. Note that $\Delta_{xy}^x = \Delta_{xy}^y$ and $\Delta_{xz}^x = \Delta_{yz}^y$ due to the symmetry relations in Eq.~\ref{sym2} that to a numerically good approximation hold even in the presence of spin-orbit coupling. From these results we see (i) the expected crossover from second order to first order at the onset of Pauli limiting and (ii) that the additional line of phase transitions is driven by a collapse of the superconducting gap on the $\alpha$ and $\beta$ Fermi sheets, with only a small feature present in $\Delta_{xy}$ arising due to the coupling induced by spin-orbit interaction. One can also note changes in curvature of the gap function before the onset of the additional first order transition, and the curious feature of a very small increase in the superconducting gap as temperature is increased through the lower critical line.
For $d_{xz}+id_{yz}$ pairing the gap function is somewhat larger and shows the expected cross over to Pauli limiting, but otherwise shows no distinguishing features. 
Fixing the temperature and varying field produces a very similar picture; for completeness we show this data in panels (e-f) of Fig.~\ref{Delta-T-B}.

\section{Discussion}

We have studied the $d$-wave pairing symmetries $d_{x^2-y^2}$ and $d_{xz}+id_{yz}$ within a realistic three dimensional tight-binding model with the electron interaction treated via two off-site Hubbard $U$ parameters, the latter fitted to reproduce the zero field superconducting transition of $1.5$K. Reassuringly, the approach then captures with excellent quantitative agreement several further experimental findings: notably the specific heat jump at zero field and the crossover temperature to Pauli limiting first order transitions at finite field. For the Knight shift suppression (the ratio of superconducting to normal state susceptibilities) we find 80\% in good agreement with the most recent experimental data \cite{chronister2020evidence}.

We find that the $d_{x^2-y^2}$ and $d_{xz}+id_{yz}$ give very similar results for the Knight shift suppression, reduction in spin moment, and zero and low field heat capacities. However at large fields, in the strongly Pauli limiting region of the H-T phase diagram, these pairing symmetries reveal strikingly different behaviour. For the chiral TRS breaking $d_{xz}+id_{yz}$ pairing we find a single phase boundary of first order transitions separating the normal and superconducting state. However for $d_{x^2-y^2}$ pairing the same region of the H-T phase diagram exhibits two phase boundaries: a line of first order transitions extending from $T=0$ and ending in a critical point at $T=0.09-0.22$K exists in addition to the upper critical field line. On crossing this additional phase boundary the material remains superconducting but suffers a significant reduction of the superconducting gap on the $\alpha$ and $\beta$ Fermi sheets, with only a minimal change on the $\gamma$ Fermi sheet. Variation of the Hubbard $U$ parameters of our model (while ensuring that the zero field critical temperature remains fixed at 1.5K) reveals these findings to be qualitatively robust: for no parameters was an additional phase boundary line in the Pauli limiting regime found for $d_{xz}+id_{yz}$ pairing, while this feature was always present for $d_{x^2-y^2}$ pairing. However, the position of the phase boundary line is sensitive to the choice of $U$ parameters.

In an earlier work exploring $p$-wave helical pairing \cite{Gupta-2020} we found a high field low temperature structure of the phase diagram different from either of the pairing symmetries explored here, namely a lower critical field line that joined the bicritical point. Taken together this suggests that the Pauli limiting region of the H-T phase diagram may represent an interesting way to distinguish pairing symmetries in $Sr_2RuO_4$. A number of previous experimental works have described features at high fields that have some resemblance to those we report here, however it is fair to say that the high field phase behaviour is not conclusively established. The calculation we present here can in future be improved by coupling the external field both with the spin (as we do here) and orbital (neglected in this work) degrees of freedom. This might be expected to bring our field values, generally too high by a factor $\sim 2$, into better agreement with experiment. Nevertheless, as the novel physics of the H-T phase diagrams we report exist in the strongly Pauli limiting region of phase space (i.e. the contribution from the spin-coupling dominates), there is reason for optimism that they will be robust to the inclusion of orbital coupling of the field.
\\
\\
\section{Acknowledgements}
This work was carried out using the computational facilities of the Advanced Computing Research Centre, University of  Bristol. J.Q., and J.A. acknowledge support from  EPSRC  through  the  project  ”Unconventional  Superconductors: New paradigms for new materials” (Grants No. EP/P00749X/1 and No. EP/P007392/1). M.G. thanks the visiting professorship program of the Centre for Dynamics and Topology at Johannes Gutenberg-University Mainz.


\end{document}